\documentclass[12pt]{article}
\usepackage{psfig,bib}
\newcommand{\avida}{{\bf Avida}}
\newcommand{\Avida}{{\bf Avida}}

\begin{document}
\mbox{}
\vskip 3cm
\centerline{\LARGE\bf \sf Selective Pressures on Genomes in Molecular Evolution}
\vskip 0.5in
\centerline{\large \bf \sf Charles Ofria$^{\dagger}$,  Christoph
  Adami$^{\star\ddagger}$, and Travis C. Collier$^\S$}
\vskip 0.2in
\centerline{$^\dagger$Center for Microbial Ecology, Michigan State University,
 East Lansing, MI 48824}
\centerline{TEL.: (517) 432-1207\ \ \  FAX: (517)353-3955} \vskip 0.2in
\centerline{ $^\star$Digital Life Laboratory 123-93} \centerline{California
Institute of Technology, Pasadena, CA 91125} \centerline{TEL.: (626)395-4516\ \ \
FAX: (626)396-0925} \vskip 0.2in \centerline{ $^\ddagger$Jet Propulsion Laboratory
MS 126-347} \centerline{California Institute of Technology, Pasadena, CA 91109}
\centerline{TEL.: (818)393-0716\ \ \ FAX: (818)393-5471} \vskip 0.2in \centerline{
$^\S$Division of Organismic Biology, Ecology, and Evolution,} \centerline
{University of California Los Angeles, Los Angeles, CA 90095} \centerline{TEL:
(310)206-4079\ \ \ FAX: (310)206-3987 }


{\noindent {\bf Running Head}: Selective Pressures on Genomes }

\vskip 0.15cm

\date{\today}

\vskip 4 cm
\newpage

\noindent \centerline{\bf Abstract}

\noindent
 We describe the evolution of macromolecules as an information transmission process and
 apply tools from Shannon information theory to it.  This allows us to
 isolate three independent, competing selective pressures that we term compression,
 transmission, and neutrality selection.  The first two affect genome length:
 the pressure to conserve resources by compressing the code, and the pressure
 to acquire additional information that improves the channel, increasing the
 rate of information transmission into each offspring.  Noisy transmission
 channels (replication with mutations) gives rise to a third pressure that acts
 on the actual encoding of information; it maximizes the fraction of
 mutations that are neutral with respect to the phenotype.  This neutrality
 selection has important implications for the evolution of evolvability.  We
 demonstrate each selective pressure in experiments with digital organisms.

\section{Introduction}

The techniques of Shannon information theory are steadily gaining application in
the field of biology, especially bioinformatics.  Information theory has been used
to study the information content of
biomolecules\nocite{gatlin72,schneider86,pavesi97,adami00} (Gatlin, 1972; Schneider
{\em et al.}, 1986; Pavesi {\em et al.}, 1997; Adami and Cerf, 2000), as well as
the mode and manner in which this information is acquired\nocite{schneider00,aoc}
(Adami {\em et al.}, 2000, Schneider, 2000), but not as a tool to isolate the
selective pressures that adapting populations are subject to. This is surprising as
information theory describes not only the difference between entropy and
information, but also the transfer and storage of information.

Recently\nocite{adami00} Adami and Cerf (2000) argued that a genomic complexity can
be defined rigorously within standard information theory as the information the
genome of an organism contains about its environment. Adami {\em et al.} (2000)
then demonstrated how this complexity must generally increase in simple evolving
systems\nocite{aoc}. Here, we focus on the selective pressures of molecular
evolution and how they contribute to the evolution of robust and complex
structures. Thus, information theory can be a tool that ties together evolutionary
biology and molecular biology.

From the point of view of information theory, it is convenient to view Darwinian
evolution on a molecular level as a collection of information transmission
channels, subject to a number of constraints.  In these channels, the organism's
genomes code for the information (a message) to be transmitted from progenitor to
offspring, and are subject to noise due to an imperfect replication process.
Information theory is concerned with analyzing the properties of such channels, how
much information can be transmitted, and how the rate of perfect information
transmission of such a channel can be maximized. This information transmission rate
can be obtained by multiplying the channel's symbol transmission rate (the rate at which symbols
can be sent from the source to the destination without taking into account the error rate) and its capacity (the maximal
information that can be encoded per symbol while being able to guarantee
faithful transmission at a given level of noise). A major assumption of the
information-theoretic treatment of evolution is that selection acts to maximize the
probability that the full genetic information is faithfully transmitted to the next
generation. This maximization involves a number of factors that we isolate below,
and guides us in elucidating the pressures evolution exerts on the genome itself.

We shall identify three pressures acting on evolving genomes and recast them in an
information-theoretic language. This allows us to formulate the forces acting
during selection from a unified point of view, while making unique predictions that
go beyond the standard population genetics treatment of evolution. The first
pressure results in the {\em compression} of a message that is not perfectly coded,
and (in the absence of any other pressures) tends to reduce message length. The
second pressure favors longer messages by increasing the amount of information
transmitted per usage of the channel. The third pressure forces populations to code
the information in a fault-tolerant manner (when mutation rates are high). While
the first two pressures are well-known within standard
theory\nocite{buerger00,baake00,baake01} (B\"urger, 2000; Baake \& Gabriel, 2000,
Baake \& Wagner, 2001), the third one is novel, and its theoretical treatment has
only been elucidated recently (van Nimwegen, 1999; Wilke {\em et al.}, 2001, Wilke,
2001b, Wilke 2001c). The pressure to code information in a fault-tolerant manner
implies that codes should evolve that are robust to deleterious mutations, or, in
other words, that the fraction of mutations that are neutral with respect to
fitness should increase.

The existence of this neutrality selection has implications for the evolution of
evolvability, an effect otherwise known as {\em
canalization}~\nocite{waddington42,mather53,gibson00}(Waddington, 1942; Mather,
1953; Gibson \& Wagner, 2000). It appears to be active in viral\nocite{burch00}
(Burch \& Chao, 2000), as well as digital (Wilke et al., 2001) populations, and may
even be responsible for the evolution of ``robustness genes" such as
Hsp90\nocite{rutherford98,wagner99,queitsch02} (Rutherford \& Lindquist, 1998;
Wagner {\em et al.}, 1999; Queitsch {\em et al.}, 2001)

\section{Selective Pressures}
Standard population genetics views the process of adaptive evolution (in the
absence of frequency-dependent selection) as the maximization of a suitably defined
fitness. Under simple conditions (a population of self-replicating molecules in a
single niche, without co-evolution) and ignoring the effects of mutation, a
genotype will dominate if its basic Malthusian parameter---the growth-rate of a
population seeded by this genotype---is highest. When analyzed from an
information-theoretic point of view, we see that the growth rate is in fact
affected by three distinct parameters: the {\it length} of the message, the maximum
{\it rate} (in units of bits/sec) at which that message can be transmitted, and the
message's robustness to mutations. We shall take a look at each one of them
separately, and demonstrate their effect on an evolving population of digital
organisms.

\subsection{Compression Selection}

No matter how information is transmitted, there is a cost for passing along each
symbol.  This general truth is obvious in biology where transmitting a symbol
involves copying a nucleotide, and this cost can be minimized simply by
transmitting fewer symbols (a shorter message). This results in a selective
pressure that we call {\em compression selection}, and illustrate it in a simple
experimental system: a population of digital organisms adapting to a virtual world
whose complexity and structure is determined by the
experimenter\nocite{adami98,lenski99,aoc,wilke02} (Adami, 1998; Lenski {\em et al},
1999; Adami {\em et al.}, 2000; Wilke \& Adami, 2002).  The experiment
demonstrating compression selection is identical in spirit to those of
Spiegelman\nocite{mills67} (Mills {\em et al.}, 1967) involving RNA replicating in
vitro.

\begin{figure}[h]
\centerline{\psfig{figure=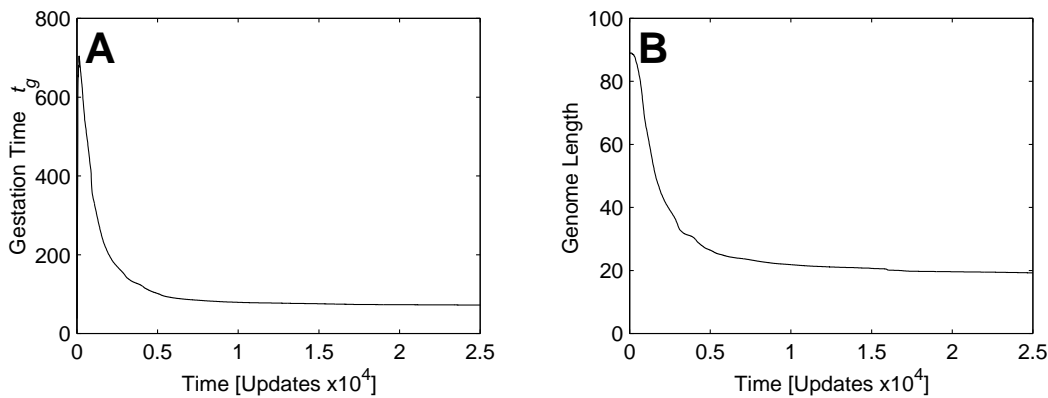,width=5.5in,angle=0}}
 \caption{Average replication time (A) and average genome length
   $\ell$ (B) for complex organisms introduced into a simple world
   (set I).
  \label{r-select}}
\end{figure}

Among a fixed-size population of self-replicators (see Methods), a single limiting
resource is allocated evenly to the organisms regardless of phenotype or genotype,
such that the only trait selected for is self-replication.  In this scenario,
compression selection will dominate the other evolutionary pressures on the
population. Since it takes time to copy each symbol, there is a pressure to shrink
the genome to the minimal length that can still self-replicate.  Taking organisms
evolved in a complex environment (see Methods) and introducing them into this
simple one, we can demonstrate this selective pressure in action.
Figure~\ref{r-select}(A) shows the adaptive drop in average replication time (CPU
time itself being the single resource in the system) as sections of the genome that
are meaningless in the simple environment are stripped away.  In
Figure~\ref{r-select}(B) we witness the corresponding drop in genome length.  Of
course, this is completely analogous to the seven-fold decrease in sequence length
observed in the evolution of RNA in a simple in vitro environment\nocite{mills67}
(Mills et al., 1967).  Naturally, the other experiments described below would be
much more difficult (if not impossible) to carry out with RNA.

Although compression selection is fundamental and unavoidable, it
is also expected to be weak in systems where the cost associated
with transmitting the message on a per-symbol basis is low.  However,
in simple molecular systems such as the one described above, as well as in
periods following the invasion of a new niche or an environmental
change, the effects on the genome can be significant due to the removal of
genes that have lost their usefulness.

\subsection{Transmission Selection}

Most resources in extant biological systems are spent building ``survival
machines'' as Dawkins (1976) eloquently put it\nocite{dawkins76}. If an organism
obtains an increased share of resources (including expected lifetime), its
opportunity for replication is correspondingly increased. This is often considered
first by biologists when thinking of interesting examples of adaptation.  In terms
of information theory, these ``survival machines'' increase the symbol transmission
rate of their channel.  Indeed, in a population of self-replicators with a single
limiting resource, an organism with twice the affinity to that resource compared to
its competitors will produce twice as many offspring, and will soon dominate.  In
other words, by taking better advantage of its environment, an organism can
increase the "bandwidth" of the channel through which it transmits its genetic
message to future generations.

\begin{table}[h]
\caption{Sets of experiments with digital organisms to study
  evolutionary pressures.
  Each set consists of 100 trials that were averaged for
  statistical significance. Second column: complexity of environment
  (S; simple, M: medium, C: complex; see Methods),
  third column: pressure for compression
  selection, fourth: constant length constraint, fifth: type of
  ancestor, sixth: mutation rate $R$.}
\vskip 0.5cm
\begin{center}
\begin{tabular}{|c|ccccc|} \hline
Set & Complexity & Compression & Constraint & Ancestor & Rate \\ \hline
I  & S & yes & no & complex & 0.75 \% \\
II   & S & no  & no & simple & 0.75 \% \\
III  & M & no  & no & simple & 0.75 \% \\
IV & C & no  & no & simple & 0.75 \% \\
V   & C & no  & yes & simple & 0.5 \% \\
VI  & C & no  & yes& simple & 1.0 \% \\
VII & C & no  & yes & simple & 1.5 \% \\
\hline
\end{tabular}
\end{center}
\end{table}

To demonstrate this selective pressure, we modified the previous environment that
favored a short sequence length, by allowing the single limiting resource to be
actively competed for.  For our digital organisms, this meant that the resource
(CPU cycles) was now distributed according to phenotype, in this case to those
phenotypes that performed selected computations (see Methods). Additionally, in
these experiments we scaled the resource allocation with genome length to
neutralize the effects of compression selection.  This allocation method ensures
that organisms with identical computational abilities are equally fit in a
mutation-free environment, regardless of genome length.  While a non-zero mutation
rate creates an additional pressure to reduce the target area for mutations, in a
complex environment we nevertheless observe a tendency for organisms to {\em
increase} sequence length in order to store information about their environment. It
is this information that modifies the organisms' phenotype, allowing them to better
exploit the environment (i.e., it increases the capacity of their channels, through
which they pass their genomic information to their offspring.)

We performed trials (sets II-IV, see Table 1) starting from a simple,
self-replicating ancestor in three environments of
differing complexity: {\em simple} where no computational tasks are
rewarded (as in our compression selection example),
{\em medium} where the completions of 10 different logical operations
are each rewarded by increased resource affinity (in turn, increasing
the channel's symbol transmission rate), and {\em complex} where
a total of 78 distinct operations are rewarded.  In this setup, we expect
that transmission selection will cause populations in more complex environments
to incorporate more information about those environments into their
genomes, resulting in relatively more fit organisms.
\begin{figure}[h]
\centerline{\psfig{figure=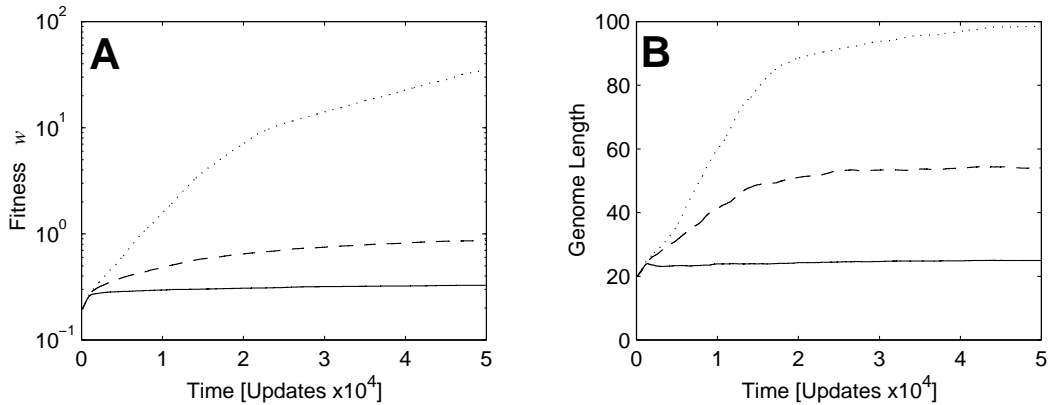,width=5.5in,angle=0}}
 \caption{Average fitness $w$ (A) and average genome length $\ell$ (B)
   displayed for three experiments at distinct levels of complexity.
   Set II (solid, simplest environment),
   set III (dashed, intermediate environment),
   and set IV (dotted, most complex world).
  \label{K-select}}
\end{figure}

Figure~\ref{K-select}(A) shows the trajectory of resource affinity divided by
replication time (the fitness $w$) as function of time.  This fitness is a
Malthusian parameter measuring rate of replication. Replication time does increase
as computational tasks are acquired, but this is more than balanced by the improved
resource affinity, resulting in an increase of the overall fitness.
Figure~\ref{K-select}(B) demonstrates that genome length increases in relation to
increasing environmental complexity, as organisms learn to exploit their
environment by acquiring information that increases the channel's symbol
transmission rate.

\subsection{Neutrality Selection}

Up to this point, we have been careful to exclude the effects of mutation.
Mutations are easily incorporated in an information-theoretical analysis as noise
affecting the information transmission channel. This inclusion reveals a third
selective pressure not easily apparent at the population level, but affecting the
rate at which complexity grows in adapting populations with implications for the
evolution of evolvability.

For illustrative purposes, suppose that all mutations are lethal, and as a
consequence directly lower the number of expected offspring by the probability of a
mutation occurring anywhere in the genome. In this case, the probability of
faithfully transmitting genetic information to the next generation, which we have
assumed is maximized by evolution, must account for the noise affecting the
transmission process.  However, just like there are many ways to express a concept
in English words, there are many messages (sequences) that contain the same meaning
(information), and thus in reality not all mutations are deleterious. If a mutation
modifies the genetic sequence without affecting the message, but does make it more
robust to further mutations, such a sequence will have an enhanced probability of
successful transmission, and therefore possesses a selective advantage. This is
most easily thought of as a {\em neutrality selection} that acts to increase the
probability that mutations are otherwise selectively neutral. While the possibility
of such a pressure has been mentioned before\nocite{clarke70,modiano81} (Clarke,
1970; Modiano {\em et al.}, 1981), its theoretical basis has only been rigorously
studied for the case that all mutations are either fully neutral or strongly
deleterious (van Nimwegen {\em et al.} 1999, Wilke 2001b). Here, we show that this
pressure is more general, and exists even when the effects of mutations are
distributed on a continuous scale.

As the replication process is not perfect, mutated genotypes often do not
contribute to the strength of a species.  Denote by $F$ (the fidelity of
replication) the probability that a birth process gives rise to a genetically
identical offspring.  If all mutations are deleterious, the true rate of growth
(fitness) of a genotype is fidelity times raw replication rate,
\begin{equation}
w_F=F w\;,
\end{equation}
which is the number of genetically identical offspring produced per unit time.  In
the language of information theory, this is the number of perfect copies of a
message received per unit time.

In reality, of course, not all mutations are detrimental. From an
information-theoretic perspective, a sequence whose information content is not
altered is as good as an un-mutated copy: the neutrality of a code is akin to a
limited error-protection. Therefore, we are interested in the probability that a
replication process gives rise to either a true copy, or one that differs from the
wild-type by neutral mutations only.

In our system (and macro-molecular evolution in general) the fidelity of the copy
process can be written in terms of the error rate per monomer copied, $R$, and the
length of the genome $\ell$, as $F=(1-R)^\ell$. Denote by $\nu$ the probability
that a mutation does not affect the replication rate, $w$, of an organism's
offspring. Then, the {\em neutral fidelity}
\begin{equation}
F_\nu=(1-R+\nu R)^\ell \label{nfidel}
\end{equation}
is the probability of either replicating accurately or else with only
neutral mutations.
Thus, the combination
\begin{equation}
w_\nu=F_\nu w \label{efit}
\end{equation}
is the replication rate of the actual genetic information (in its original or a
neutrally modified form).  Because $F_\nu$ is an increasing function of $\nu$, a
pressure to increase $w_\nu$ implies that there is a selective pressure maximizing
$\nu$.  This is neutrality selection. Note, however, that evolution acts to
increase the average $w_\nu$ in the population (van Nimwegen {\em et al.} 1999;
Wilke 2001c). Thus, $\nu$-selection implies that selection at high mutation rates
acts on groups of mutually neutral genotypes, rather than on the individuals
themselves.

Before discussing evidence for neutrality selection in experiments with digital
organisms, let us point out that the above formulae can be recast in terms of
standard population biology of clonal or randomly mating sexual populations, if $w$
is understood as the fitness contribution of a locus or gene.

Equilibrium (meaning mutation-selection balance) demands that the true rate of
growth of the wild-type $Fw_0$ equals the average growth rate $\bar w$, or, if
there are neutral mutations, $F_\nu w_0=\bar w$. (Note that the neutrality $\nu$
going into the calculation of $F_\nu$ is the average neutrality of a population
grown from the wild-type (Wilke 2001c), which we approximate in the experiments
below by the average neutrality of all one point mutants of the wild-type, see
Methods).

We can then calculate the {\em mutational load} on the population (the fraction of
the population that dies each generation because of deleterious mutations) as
\begin{equation}
L=1-\frac{\bar w}{w_0}=1-F_\nu\;.
\end{equation}
Now, using Eq.~(\ref{nfidel}) and the fact that $R << 1$, we obtain
\begin{equation}
L\approx 1- e^{-R\ell(1-\nu)}\;. \label{load}
\end{equation}
We see that a population can reduce its mutational load, at constant mutation rate,
by increasing the neutrality (fraction of all possible one-point mutations that are
not deleterious to the organism) $\nu$ of its genomes. This has been noted
independently by van Nimwegen {\em et al.} (1999) and by Wilke (2001b) for the case
of strictly neutral evolution.

A decrease of mutation rate (e.g., by the development of error correction)
would provide another means of lowering the mutational load.  However, a
lower mutation rate would also result in reduced genetic variability.
According to Fisher's Fundamental Theorem, a lower genetic variability
correlates to lower adaptive potential.  As a consequence, neutrality
selection may be thought of as an indirect selection for evolvability.

We demonstrate the influence of neutrality selection using \Avida\ by directly
measuring the neutrality of the organisms, $\nu$, (see Methods) as a function of
mutation rate (i.e., for populations adapting under different fixed mutation rate).
We examined three sets of trials subjected to different mutation rates (sets
V-VII), all in identical complex environments (equalizing the pressure of
transmission selection) and constrained to remain at a genome length of 100
instructions (preventing length changes, and hence nullifying compression
selection, see Table 1).

\begin{figure}[h]
\centerline{\psfig{figure=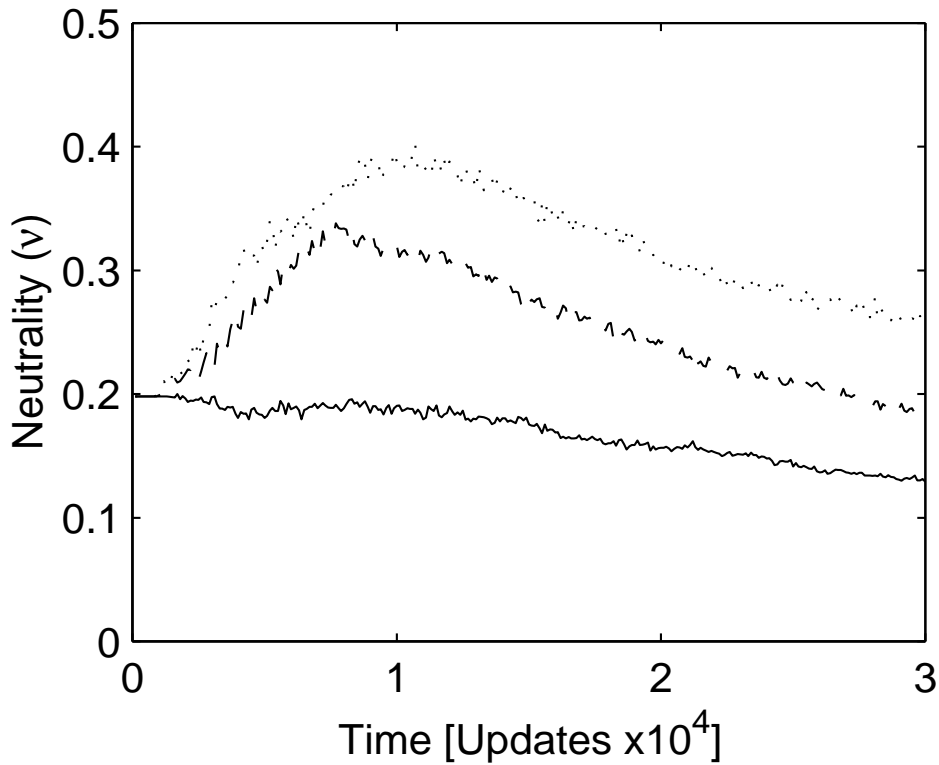,width=5in,angle=0}}
 \caption{Average neutrality for sets V (solid), VI (dashed),
  and VII (dotted), with fixed genome length in a complex environment
  (see Table 1).
  Their per-copy mutation rates are 0.5\%, 1.0\%, and 1.5\% respectively.
  \label{nu-select}}
\end{figure}

Higher mutation rates result in a stronger pressure to increase neutrality. This
trend is observed in the measured neutralities of populations adapting in
environments at different mutation rates, see Figure~\ref{nu-select}. Additionally,
the interplay between transmission and neutrality selection is apparent from the
slow decrease in neutrality for all three sets due to information that increases
resource affinity, and is incorporated into the genomes.  Since length is fixed,
the genome sequence is being ``filled up'' with information in the form of adaptive
mutations that become fixed in the genome, and neutrality must decrease. In other
words, as the population gains information about the environment, sites that were
once neutral will now cause a loss of information were they to be mutated.

\section{Methods}
Evolutionary experiments were performed on digital
organisms\nocite{adami98,lenski99,aoc} (Adami, 1998; Lenski {\em et al.}, 1999;
Adami {\em et al.}, 2000) implemented with version 1.3 of the {\avida} software.
This software and its documentation can be obtained from {\tt
 http://dllab.caltech.edu/avida}.  The {\avida} system controls
populations of self-replicating computer programs in a complex and
noisy environment within a computer's memory. Self-replicating
programs may be viewed, within limits, as the computational analogs
of catalytically active RNA sequences that serve as the template of
their own self-reproduction.  Populations are propagated for 50,000
updates (an arbitrary unit of time: every update represents the
execution of an average of 30 instructions per individual in the
population) from the same single ancestor genotype of length 20
instructions (except for set I, see Table 1) to the carrying capacity
of 3,600 individuals. Sequence length is variable except for trials
V-VII. Mutations occur during replication, at the rate of $R$
mutations per instruction copied per organism. Default settings were
used in all experiments unless otherwise indicated.

Computations can be carried out by evolved programs if they develop sequences of
code (``genes'') that perform logical, bitwise, computations on random numbers
provided in their environment.  Such genes evolve spontaneously if the performance
is rewarded with bonus CPU cycles (their unit of energy).  The complexity of the
environment can be controlled by changing the number of logical operations whose
performance is rewarded. The development of the necessary code to perform such
computations is, in the digital world, the analog of the evolution of a sequence or
gene enabling the organism to catalyze exothermic reactions leading to faster
replication.  Outside materials (in this case binary numbers) are taken in and
processed for an energy advantage to the organism (see Wilke \& Adami, 2002).

From an information theoretic viewpoint, the rewards for these computations
directly increase the symbol transmission rate of the channel, because longer messages
can be sent per usage of the channel. In our simplest world (set II), no such
computations are rewarded, and thus no genes beyond replication develop. In the
intermediate world (set III), 10 computations are rewarded, and different
combinations of these functions are acquired in the different replicas of that set.
In the third and most complex set, 78 different operations are potentially
rewarded.

Neutrality was measured by obtaining the fitness of all possible one-point mutants
of the most abundant genotype, which allowed us to classify mutations as either
deleterious, neutral, or beneficial. Neutral mutations are those for which the
fitness has changed less than $1/N$, where $N$ is the population size (3600
organisms, for all experiments presented here).  The neutrality is the fraction of
all mutations that are either neutral or beneficial (as these are those that do not
decrease the information content).

\section{Conclusion}

In natural systems, neutral mutations often occur on a non-critical nucleotide
(synonymous substitutions), or else in code that is never expressed.  That most
molecular evolutionary change is neutral with respect to the phenotype has been
widely accepted since Kimura's seminal work on the subject\nocite{kimura83}
(Kimura, 1983). Explaining why so much evolutionary change is neutral has been
difficult, and has complicated the discussion of increasing genomic complexity
(even though genome length has generally been increasing over the course of
evolutionary history\nocite{cavalier85}, Cavalier-Smith, 1985), since the
correlation between genome length and organismic complexity appears weak---the
so-called C-value paradox.

Here, we point out that we can think of the quantity maximized by the process of
evolution as the probability of faithfully passing on the information encoded in
the genome on to the next generation, which is simply the Malthusian parameter
($w$) multiplied by the neutral fidelity (\ref{nfidel}) of the organism. This
implies that a population can minimize its mutational load by increasing the
probability that a mutation is neutral, or, equivalently, decreasing the
probability that a mutation is deleterious.  Experiments with digital organisms
confirm this view. In particular, an analysis of an aggregate of approximately 10
million generations of the evolution of digital organisms has never revealed a
single case in which the effective fitness, Eq.~(\ref{efit}), has decreased.

The selection of neutrality has a number of interesting consequences when
translated to population biology.  First, it might provide a convincing mechanism
to explain the actual amount of neutrality in the genomes of higher organisms.
Furthermore, because a higher neutrality implies both larger fitness variations and
the potential for more information acquisition, neutrality selection implies a
selective pressure for the evolution of evolvability (see, e.g.,
\nocite{wagner96,gerhart97,ancel00}Wagner \& Altenberg, 1996; Gerhart \& Kirschner,
1997, and Ancel \& Fontana, 2000) that is molecular at its origin, and provides
added support for the idea that robustness is actively selected for in the
evolution of development\nocite{gibson00} (Gibson \& Wagner, 2000). That it may
also have left traces in the genomes of higher organisms in the form of proteins
that confer neutrality to other proteins (Rutherford \& Lindquist, 1998; Wagner
{\em et al.}, 1999;  Queitsch {\em et al.}, 2000) is a tantalizing suggestion that
awaits confirmation.

\vskip 0.25cm \noindent{Acknowledgements} \vskip 0.25cm This work was supported by
the National Science Foundation under grants PHY-9723972 and DEB-9981397. Access to
a Beowulf system was provided by the Center for Advanced Computation Research at
the California Institute of Technology. Part of this work was performed at the Jet
Propulsion Laboratory, California Institute of Technology, under contract with the
National Aeronautics and Space Administration. We thank Richard Lenski and Claus
Wilke for valuable suggestions.

\end{document}